# MAPPING OF TRANSRECTAL ULTRASOUND PROSTATE BIOPSIES : QUALITY CONTROL AND LEARNING CURVE ASSESSMENT BY IMAGE PROCESSING

**Authors :**

Mozer, Pierre [1,2] MD, PhD
Baumann, Michael [1] PhD
Chevreau, Gregoire [1] PhD
Moreau-Gaudry, Alexandre [2,3] MD, PhD
Bart, Stephane [1] MD
Renard-Penna, Raphaele [4] MD
Comperat, Eva [5] MD
Conort, Pierre [1] MD
Bitker, Marc-Olivier [1] MD
Chartier-Kastler, Emmanuel [1] MD, PhD
Richard, Francois [1] MD
Troccaz, Jocelyne [2] PhD

**Affiliations :**

1 : APHP - Pitié-Salpétriere, Urology;
2 : Institut de l'Ingénierie et de l'Information en Santé, GMCAO, TIMC-IMAG
3 : CIC-IT, Grenoble University Hospital, Département de Méthodologie
4 : APHP - Pitié-Salpétriere, Radiology
5 : APHP - Pitié-Salpétriere, Department of Pathology

**Address of corresponding author :**
Urology department
Groupe Hospitalier Pitié-Salpétrière
91, Bd de l'Hôpital
75013 Paris, France
Phone : +33 1 42 17 71 00
Email : Pierre.mozer@psl.aphp.fr

**Keywords:**
prostate biopsy;
3D ultrasound;
image based tracking; image processing;
quality control;
learning curve

**Objective:**

TRUS (transrectal ultrasound) mapping prostate biopsies is of fundamental importance for either diagnostic purposes or the management and treatment of prostate cancer but the localization of the cores seems inaccurate. Our objective was to evaluate the capacities of an operator to plan transrectal prostate biopsies under 2D TRUS guidance using a registration algorithm to represent the localisation of biopsies in a reference 3D ultrasonography volume.

**Material and Methods:**

32 patients underwent a series of 12 prostate biopsies under local anaesthesia performed by one operator using a TRUS probe combined with specific third party software, to verify that the biopsies were indeed conducted within the planned targets.

**Results :**

The operator reached 71% of the planned targets with substantial variability that depended on their localisation (100% success rate for targets in middle and right parasagittal parts *vs*. 53% for targets in the left lateral base). Feedback from this system after each series of biopsies enabled the operator to significantly improve his dexterity over the course of time (first 16 patients: median score = 7/10 and cumulated median biopsy length in targets of 90 mm vs. a score of 9/10 and a cumulated median length of 121 mm for the last 16).

**Conclusion :**

In addition to being a useful tool to improve the distribution of prostate biopsies, the potential of this system is above all the preparation of a detailed "map" of each patient showing biopsy zones without significant change in routine clinical practices.

**Introduction and Objectives:**

Mapping transrectal ultrasound (TRUS) prostate biopsies is of fundamental importance for either diagnostic purposes or the management and treatment of prostate cancer. At the present time, there is room for improvement in their transrectal distribution, since the correlation is low between biopsy results and the pathology examination of radical prostatectomy specimens (1). This has led some operators to conduct transperineal biopsies, using a template. This procedure is difficult in routine clinical use and its accuracy is subject to caution. It is in fact well known that the prostate has significant mobility (3; 2) and the ultrasonographic planning systems do not take into account those movements at the time of puncture (4).

Modern 3D-TRUS probes allow acquiring high-quality volumes of the prostate in a few seconds. We have developed a process to track the prostate in 3D TRUS images. With a reference volume acquired at the beginning of the session and another during each biopsy, it is possible to rapidly determine the spatial relationship between the reference prostate and every biopsy volume by automatically aligning images. Thus, without changing usual clinical practices of endorectal biopsies, in a few seconds this system provides accurate biopsy cores localization by representing them in a single 3D ultrasound volume. The process is implemented on a computer system placed aside the ultrasound scanner during the procedure.

Our main goal was determining an operator's ability to conduct a predefined planning of 12 biopsies using the most widespread technique involving biopsy guidance with a 2D transverse TRUS image. We used our registration process to quantify the sampling of the gland, by measuring: first, the rate of biopsies successfully reaching predefined targets; second, the length of the biopsy cores removed from the same predefined targets. In order to quantify the feedback contribution of the system, the operator's learning curve was established.

**Material and Methods:**

*Patients and operator:*

After approval by an ethics committee, one operator performed 12-core TRUS biopsies on 32 patients from November 2006 to March 2007, according to a classical 12-core protocol. Informed consent was obtained from all subjects. Patients were supine on a table with stirrups. Biopsies were performed using transverse 2D tracking provided by a 3D TRUS probe (RIC5-9 on a Voluson-i, both from GE Healthcare) and standard biopsy needles (18 G, Biopsy, Bard, Covington, GA, USA).

The operator is a right-handed urologist, assistant professor, having conducted more than 100 prostate biopsies before the present study.

The patient inclusion criterion was PSA > 4.0 ng/mL: the median PSA was 8 ng/mL (range 4 to 18). All patients received ciprofloxacin two hours before the biopsies. Average prostate volume was 45 mL (range 20 to 100). The biopsy was conducted after local anaesthesia with 10 cm$^3$ (5 cm$^3$ on each side) of 2% lidocaine injected around the periprostatic nerve plexus under conventional 2D TRUS guidance.

*Biopsies: intra-operative acquisitions*

Before the first biopsy, a 3D reference volume including the prostate was acquired. The probe was then switched to 2D mode for biopsy targeting. After each biopsy gunshot, the needle was left indwelling in the prostate for an average of 3 seconds, as a 3D TRUS volume was acquired. During this acquisition, the operator paid attention to apply minimal force on the probe to minimise deformation of the prostate.

*Biopsies: post-operative registration and measurements*

The needle was delineated in each biopsy volume and was automatically mapped into the reference volume with a rigid image-based registration algorithm (5). This algorithm is fully automatic and is based on the statistical analysis of the composition of images. This organ-based tracking process enables both the movements of the TRUS probe and those of the prostate to be followed. In order to validate registration, we pointed out clearly visible point-like fiducials, e.g. calcifications in the prostate. The distances between corresponding fiducials after application of the registration transform were used as the gold standard to determine accuracy.

After registration, the biopsy cores can be represented in the reference volume (Figure1) enabling their spatial distribution to be analysed. Quantitative analyses were conducted by reslicing the reference volume in the coronal plane and the 12 pre-operative square target areas were superimposed on these images. Target areas are named with the following convention: B=Base, M=Mid-Gland, A=Apex, L=Lateral, P=Parasagittal, R=Right and L=Left. For example, the right parasagittal base area is noted BPR.

For each square coronal target area mentioned above, we computed the rate of biopsies hitting the target and the biopsy length inside the target. The length measured was that of the cylinder removed from the target by the needle, i.e. included in a 20 mm long zone starting 5 mm from the needle tip. When a biopsy overlapped several targets, the two greatest lengths were recorded. A biopsy was considered successful when its length within the targeted area was more than 3 mm due to the accuracy of the algorithm reported previously (5).

Since the apical lateral target areas are small on pre-operative planning, we merged them with the apical parasagittal targets for analysis. This merging was conducted by averaging biopsy tissue lengths in the lateral and parasagittal apical areas.

Thus, for each patient, 10 «targets» were analyzed to study the sampling procedure and determine the operator's capacity to carry out a planning.

The operator was made aware of the results several days after biopsies and was able to see on a 3D interface were biopsies were performed.

*Statistical analysis*

In terms of operator capacities to conduct a predefined planning, a quantitative analysis of the distribution of biopsied tissue lengths was conducted using Friedman's non-parametric test. The effect of prostate volume on the length of samples recovered from targets was determined using a linear regression.

The learning curve was analysed by separating patients into two groups according to chronological order: the first 16 patients were in group 1 and the last 16 in group 2. The operator aimed at 10 targets and a score from 0 to 10 was calculated for each patient. The Mann-Whitney non-parametric test was used to compare the score distribution between the two groups.

The same methodology was applied to the analysis of sampling in the biopsied areas.

The threshold of significance adopted for this study was 0.05%.

**Results:**

No post-biopsy complication was reported. Average pain was scored as 2 (range 1 to 4) on a scale from 0 to 10.

*Analysis of robustness and accuracy of image registration*

The accuracy of the registration method was validated on 237 3D images acquired during biopsies with an average error < 2 mm and a maximum error < 4 mm. Registration between two volumes was computed on average in 6 seconds. The success rate of all the registrations was 97% (375 good registrations on 384 volumes). The registration system failed 9 times, but these failures resulted from images containing artefacts caused by the presence of air between the probe and tissues.

*Analysis of the operator's capacities to reach planning targets*

All results are summarised in Table 1. The operator reached his target on average in 71% of the cases. The success rate was maximal in the MPR part (100%) and minimal in BLL (53%). Overall, the ratio was maximal in the centre and decreased as the planning approached the boundaries of the prostate.

The distribution of biopsied tissue lengths was significantly dependent in the targeted area ($p = 1.54^{-10}$). It may be noted that the length of tissue sampled from the apex and the BLL were less than half that expected. Finally, no effect of prostate volume on these lengths could be shown ($p = 0.5$).

*Analysis of the operator's learning curve in reaching planning targets*

For group 1, the operator reached the target in 66% of cases compared to 77% in group 2. The median score in group 1 is 7 (min = 4, max = 9) and is significantly lower ($p = 0.046$) than in group 2 (median = 9, min = 2, max = 10).

The median cumulated lengths of biopsies from inside the targets was 90mm (min=37, max=130) for group 1 and 121mm (min = 32, max = 156) for group 2. The size of the sample is statistically significant (p = 0.042).

*Analysis of prostate biopsy sampling*

The length of tissues sampled was longest in the MPR zone (median = 30mm) and shortest in the right apex (median = 7mm). The distribution of lengths biopsied significantly depended on the area (p= $5^{-16}$). Parasagittal zones were sampled more than lateral zones.

*Analysis of the learning curve in prostate biopsy sampling*

The median cumulated length of biopsies in parasagittal zones was 106mm (min = 43, max = 165) for the first 16 patients and 81mm (min = 51, max = 140) for the last 16 patients. The size reduction of samples taken from these zones was statistically significant (p = 0.012). The median cumulated length of biopsies in lateral zones was 49mm (min = 12, max = 103) for the first 16 patients and 79mm (min = 20, max = 124) for the last 16 patients. The increase in length sampled from these zones is statistically significant (p = 0.044). The zones sampled thus shifted over time from the parasagittal part to the lateral part of the prostate.

**Discussion**

*Accuracy, robustness and integration of the system*

One of the advantages of this system is that each biopsy is based on images of the prostate with the biopsy needle in place and not on the relative position of the puncture needle with respect to the ultrasound images acquired several minutes beforehand. This is an important finding since, in contrast to other devices such as TargetScan® used for transrectal biopsies (6) or a brachytherapy template used for perineal biopsies, our system incorporates the movements of the gland. It is important to note that this approach does not rely on any external optical or magnetic tracking system or passive arm holding the probe, which means that no cumbersome additional material is required since the method involves uniquely the analysis of ultrasound images.

The major drawback of the algorithm used in this study is that it does not yet integrate deformations of the prostate caused by pressure from the endorectal probe. This is why the operator paid special attention to avoid deforming the gland during acquisition of images; in addition, evaluation of the accuracy of the algorithm verified that there was little or no deformation of the prostate in contact with the probe. Mean system accuracy was determined to be 2 mm with a maximal error of 4 mm in the three spatial dimensions and in our opinion has no clinical relevance.

To our knowledge, this is the first system that enables the localization of endorectal prostate biopsies with a real 3D quantified accuracy.

By increasing the time of biopsies by only a few minutes, this technique leads to a quality control of biopsy localizations specifically for each patient. System robustness should nevertheless be verified by a study involving several operators.

*Operator accuracy, study of sampling and the learning curve*

The operator in the present study had previously participated in a work using a phantom, comparing the distribution of biopsies conducted by 14 different operators (7). No significant difference was found between the operator in the former study and the 13 other clinicians. It can thus be considered that this operator is representative.

The success rate for reaching a target that the operator envisions mentally is relatively low (mean 71%) and indicates the difficulties in transferring 3D coordinates while being guided only by a 2D image. Feedback given by the system in the course of this work enabled the operator to significantly improve the distribution of prostate biopsies over time. Nevertheless, the patient population in this study was too small to prove that an increase in the number of positive biopsies resulted from better targeting. In the same vein, the capacities of different operators should be compared in order to confirm these results.

*Potentials*

There are a number of potentials for this technique. It is theoretically possible to merge series of biopsies done at different times provided the volume or shape of the prostate does not vary significantly. This would, for example, verify that a suspicious zone on prior biopsies was again sampled, rather than an adjacent region. Similarly, in the context of active surveillance requiring repeated biopsies, the relative position of each biopsy at different times would increase the confidence level of this therapeutic approach. Finally, in the case of repeated biopsies resulting from a strong suspicion of cancer, it would ensure that different zones were effectively punctured.

Currently, due to the time needed to acquire a 3D TRUS volume, the system does not allow a real-time feedback. Nevertheless, 3D TRUS imaging systems in research laboratories provide about 5 volumes per second and this algorithm should be fast enough to achieve frame-rates of 5Hz by taking advantage of the massive parallelisation capacities provided by modern high-end computers enabling continuous full 3D image-based tracking. It could be possible to select a target in the reference volume and guide the clinician to reach it. Moreover, work is in progress to take into account both deformations and merged pre-operative MRI and 3D TRUS reference volumes to select the target in MRI images and track it with high accuracy in this context.

**Conclusion**

This study has evaluated the capacities of an operator to reproduce a predefined planning and also the real sampling conducted during transrectal biopsies under 2D ultrasonography guidance. In the entire patient population, the operator reached only 71% of the designated targets, with a substantial variability depending on their localization. These localization errors are the cause of an under-sampling of the peripheral part of the gland and an over-sampling of the parasagittal part. Feedback after the procedure significantly improved the rate of targets reached and thus improved tissue sampling within the gland.

In addition to being very useful for improving prostate biopsies, the potential of this system is above all the possibility of creating a precise map of zones sampled for each patient. This biopsy localization data can be used not only to make a therapeutic decision, but also to be registered for possible near-future merging with biopsies conducted at different times in the course of surveillance.

***Acknowledgements :***

Our thanks to the French Association of Urology (AFU), Agency to Advance French Research (ANR), Assistance Publique - Hôpitaux de Paris (AP-HP) and Koelis Augmented Surgery.

## *References:*

1. Salomon L, Colombel M, Patard JJ, Lefrere-Belda MA, Bellot J, Chopin D, et al. Value of ultrasound-guided systematic sextant biopsies in prostate tumor mapping. Eur Urol. 1999;35:289-293.

2. Artignan X, Rastkhah M, Balosso J, Fourneret P, Gilliot O, Bolla M. [Quantification of prostate movements during radiotherapy]. Cancer Radiother. 2006 ;10:381-388.

3. Padhani AR, Khoo VS, Suckling J, Husband JE, Leach MO, Dearnaley DP. Evaluating the effect of rectal distension and rectal movement on prostate gland position using cine MRI. International Journal of Radiation Oncology*Biology*Physics. 1999;44:525-533.

4. Lagerburg V, Moerland MA, Lagendijk JJ, Battermann JJ. Measurement of prostate rotation during insertion of needles for brachytherapy. Radiotherapy and Oncology. 2005 ;77:318-323.

5. Baumann M, Mozer P, Daanen V, Troccaz J. Towards 3D ultrasound image based soft tissue tracking: a transrectal ultrasound prostate image alignment system. Med Image Comput Assist Interv. 2007 ;10(Pt 2):26-33.

6. Andriole GL, Bullock TL, Belani JS, Traxel E, Yan Y, Bostwick DG, and al. Is there a better way to biopsy the prostate? Prospects for a novel transrectal systematic biopsy approach. Urology. 2007;70(6 Suppl):22-28.

7. Long J, Daanen V, Moreau-Gaudry A, Troccaz J, Rambeaud J, Descotes J. Prostate Biopsies Guided by Three-Dimensional Real-Time (4-D) Transrectal Ultrasonography on a Phantom: Comparative Study versus Two-Dimensional Transrectal Ultrasound-Guided Biopsies. Eur Urol. 2007;52:1097-1104.

| Target | | # of biopsies | All patients % (#) of biopsies inside the target | Biopsy length inside the target (mm) Median (First – Third) Quartile | # of biopsies | First 16 patients % (#) of biopsies inside the target | Biopsy length inside the target (mm) Median (First – Third) Quartile | # of biopsies | Last 16 patients % (#) of biopsies inside the target | Biopsy length inside the target (mm) Median (First – Third) Quartile |
|---|---|---|---|---|---|---|---|---|---|---|
| Base Lateral (BL) | Right | 32 | 72% (23) | 15 (12-17) | 16 | 62% (10) | 15 (13-17) | 16 | 81% (13) | 15 (12-17) |
| | Left | 32 | 53% (17) | 12 (8-15) | 16 | 44% (7) | 8 (6-14) | 16 | 62% (10) | 15 (10-16) |
| Base Parasagittal (BP) | Right | 32 | 62% (20) | 16 (14-20) | 16 | 50% (8) | 15 (7-18) | 16 | 75% (12) | 17 (16-20) |
| | Left | 32 | 66% (21) | 15 (11-20) | 16 | 62% (10) | 15 (14-18) | 16 | 69% (11) | 16 (11-20) |
| Mid Lateral (ML) | Right | 32 | 81% (26) | 15 (12-17) | 16 | 75% (12) | 15 (12-18) | 16 | 88% (14) | 15 (12-18) |
| | Left | 32 | 72% (23) | 15 (13-17) | 16 | 81% (13) | 14 (13-14) | 16 | 62% (10) | 18 (16-19) |
| Mid Parasagittal (MP) | Right | 32 | 100% (32) | 16 (12-18) | 16 | 100% (16) | 17 (13-19) | 16 | 100% (16) | 15 (13-17) |
| | Left | 32 | 88% (28) | 16 (14-18) | 16 | 92% (15) | 16 (14-19) | 16 | 81% (13) | 18 (14-18) |
| Apex (Lateral + Parasagittal) | Right | 60 | 72% (43) | 7 (5-10) | 30 | 50% (15) | 7 (5-12) | 30 | 94% (28) | 7 (5-9) |
| | Left | 59 | 59% (35) | 8 (6-13) | 29 | 62% (18) | 7 (5-8) | 30 | 56% (17) | 12 (8-15) |
| Sum/Average | | 375 | 71% (268) | | 187 | 66% (124) | | 188 | 77% (144) | |

**Table 1:** Targeting accuracy evaluation results for the 32 patients

*Figure1:*
Examples of biopsy distributions in the coronal plane for 2 different patients. Left : Patient of the first group – Left base is not sampled. Right : Patient of the second group - Good sampling.

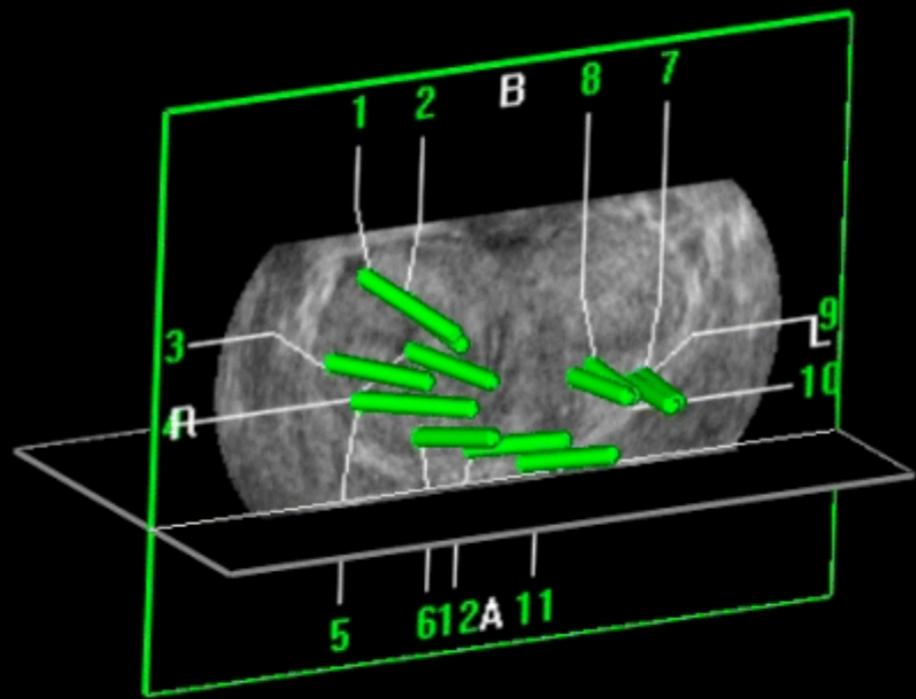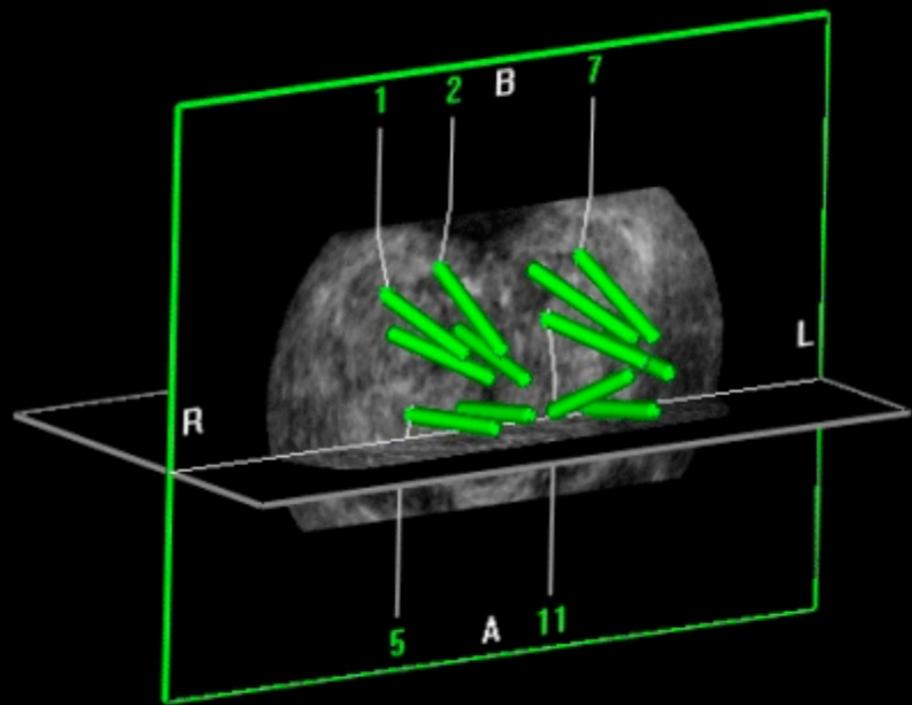